\documentclass[aps,prd,preprint,tightenlines,
   showpacs,nofootinbib]{revtex4}
\newcommand{\PRE}[1]{{#1}}   

\usepackage{bm}
\usepackage{epsfig}

\newcommand{\postscript}[2]{\setlength{\epsfxsize}{#2\hsize}
   \centerline{\epsfbox{#1}}}
\newcommand{\mweak}{M_{\text{Weak}}}
\newcommand{\mplanck}{M_{\text{Pl}}}

\newcommand{\ev}{\text{eV}}

\newcommand{\gev}{\text{GeV}}
\newcommand{\tev}{\text{TeV}}

\newcommand{\cm}{\text{cm}}

\newcommand{\s}{\text{s}}

\newcommand{\eg}{{\em e.g.}}

\newcommand{\eqref}[1]{Eq.~(\ref{#1})}
\newcommand{\eqsref}[2]{Eqs.~(\ref{#1}) and (\ref{#2})}
\newcommand{\secref}[1]{Sec.~\ref{#1}}

\newcommand{\mKK}{m_{\text{KK}}}
\newcommand{\mSUSY}{m_{\text{SUSY}}}
\newcommand{\TRH}{T_{\text{RH}}}

\begin{document}

\preprint{UCI-TR-2003-27}

\title{
\PRE{\vspace*{1.5in}}
Graviton Cosmology in Universal Extra Dimensions
\PRE{\vspace*{0.3in}}
}

\author{Jonathan L.~Feng}
\affiliation{Department of Physics and Astronomy,
University of California, Irvine, CA 92697, USA
\PRE{\vspace*{.3in}}
}
\author{Arvind Rajaraman}
\affiliation{Department of Physics and Astronomy,
University of California, Irvine, CA 92697, USA
\PRE{\vspace*{.3in}}
}

\author{Fumihiro Takayama%
\PRE{\vspace*{.2in}}
}
\affiliation{Department of Physics and Astronomy,
University of California, Irvine, CA 92697, USA
\PRE{\vspace*{.3in}}
}


\begin{abstract}
\PRE{\vspace*{.1in}} In models of universal extra dimensions, gravity
and all standard model fields propagate in the extra
dimensions. Previous studies of such models have concentrated on the
Kaluza-Klein (KK) partners of standard model particles.  Here we
determine the properties of the KK gravitons and explore their
cosmological implications. We find the lifetimes of decays to KK
gravitons, of relevance for the viability of KK gravitons as dark
matter. We then discuss the primordial production of KK gravitons
after reheating.  The existence of a tower of KK graviton states makes
such production extremely efficient: for reheat temperature $\TRH$ and
$d$ extra dimensions, the energy density stored in gravitons scales as
$\TRH^{2+3d/2}$.  Overclosure and Big Bang nucleosynthesis therefore
stringently constrain $\TRH$ in all universal extra dimension
scenarios.  At the same time, there is a window of reheat temperatures
low enough to avoid these constraints and high enough to generate the
desired thermal relic density for KK WIMP and superWIMP dark matter.
\end{abstract}

\pacs{11.10.Kk, 12.60.-i, 95.35.+d, 98.80.Cq}

\maketitle

\section{Introduction}
\label{sec:introduction}

Since the initial work of Kaluza and Klein~\cite{KK}, models of
extra dimensions have played an important part in particle theory.
These ideas have been revived in models of $\tev$-scale universal
extra dimensions
(UED)~\cite{Antoniadis:1990ew,Lykken:1996fj,Appelquist:2000nn}. In
these models, as in Kaluza-Klein (KK) theory, both gravity and
standard model (SM) fields propagate in $D=4+d$ dimensions.

In models where the extra dimensions are tori, every known particle
has a tower of KK partner particles, each carrying KK number.
Momentum conservation in the extra dimensions implies KK number
conservation.  Such models predict a plethora of unseen massless
particles and a tower of stable KK particles, both of which are
phenomenologically problematic.

For more general geometries, unwanted massless modes may be projected
out, and there are fewer or no stable KK particles.  A particularly
interesting situation occurs when there is only a $Z_2$ symmetry
(called KK-parity) in the extra dimensions. An example of this
situation is the case $D=5$ where the extra dimension is the orbifold
$S^1/Z_2$, where $S^1$ is a circle of radius
$R$~\cite{Appelquist:2000nn}. For SM fields, the geometry projects out
half of the massless modes, leaving only the 4D SM degrees of freedom,
and also breaks KK number conservation, rendering almost all KK
partners unstable.  However, the geometry preserves KK-parity, and so
the lightest KK particle (LKP) is stable, making it a viable
realization of dark matter with an extra-dimensional
origin~\cite{Kolb:fm}. Among SM KK partners, the LKP is often $B^1$,
the KK partner of the hypercharge boson~\cite{Cheng:2002iz}.  For
$\mKK \sim \tev$, $B^1$ is an excellent weakly-interacting massive
particle (WIMP) dark matter candidate, with a thermal relic density
consistent with observations and promising prospects for
detection~\cite{Servant:2002aq,Cheng:2002ej,Hooper:2002gs,%
Majumdar:2002mw,Servant:2002hb,Bertone:2002ms,Bringmann:2003sz}.

These studies, as well as those exploring the
collider~\cite{Cheng:2002iz,Rizzo:2001sd,Macesanu:2002db,%
Petriello:2002uu,Cheng:2002rn,Oliver:2002up} and
low-energy~\cite{Appelquist:2000nn,Agashe:2001ra,Appelquist:2001jz,%
Buras:2002ej,Appelquist:2002wb,Chakraverty:2002qk,Mohapatra:2002ug,%
Buras:2003mk} implications of UED, have focused on the SM KK spectrum,
neglecting the gravitational sector.  KK gravitons couple extremely
weakly, so their effects are insignificant for colliders and
low-energy experiments.  Cosmologically, however, KK gravitons may
play an important role.  For example, the LKP may be the lightest KK
graviton~\cite{Feng:2003xh,Feng:2003uy}. It is not easy to verify
this, since loop diagrams involving gravitons are divergent. On
general grounds, however, one expects the mass corrections to be
$\delta m / m \sim (\mplanck R)^{-2}$, since gravitons decouple as
$\mplanck \to \infty$.  This is potentially very small, and, since
loop corrections to other KK particle masses are typically
positive~\cite{Cheng:2002iz}, it is not unlikely that in UED theories,
the lightest KK graviton is the LKP.  In this case, all SM KK
particles eventually decay to the KK graviton, and the KK graviton is
the only possible KK dark matter candidate.

KK graviton dark matter interacts only gravitationally and is
superweakly-interacting massive particle, or superWIMP, dark
matter~\cite{Feng:2003xh,Feng:2003uy}. As such, it is impossible
to detect in conventional dark matter experiments.  However, its
potential impact on Big Bang nucleosynthesis (BBN), the cosmic
microwave background, and the diffuse photon flux provides
alternatives for dark matter searches.  In evaluating such
signals, it is important to have accurate, rather than just order
of magnitude, results for decay times of WIMPs to superWIMPs.  We
determine these in this study.

In addition, KK gravitons may be produced in the early universe during
the reheating era following inflation.  The phenomenon of gravitino
production after reheating in supersymmetric scenarios is
well-studied~\cite{Pagels:ke,Bolz:1998ek}.  The case of UED is
qualitatively different, however, as there is an infinite tower of new
particle states that may be populated at high temperatures, with the
density of states growing rapidly at large masses, especially for
large $d$.  As we will see, in UED primordial KK graviton production
is in fact very efficient, and constraints on dark matter abundances
and BBN provide stringent bounds on early universe cosmology in all
UED scenarios, irrespective of which particle is the LKP and other
spectrum details.

The paper is organized as follows. In \secref{sec:torus} we derive the
interactions of KK gravitons coupled to UED SM fields in general
toroidal dimensions. In \secref{sec:orbifold} we then obtain the KK
graviton interactions for UED orbifold compactifications, specializing
to the case of the $S^1/Z_2$ orbifold discussed above.  For this case,
we first show that boundary terms coming from the ends of the interval
are typically negligible and then determine the spectrum and bulk
interactions of the orbifold theory.  In \secref{sec:widths}, we find
the decay rates for next-to-lightest KK particle (NLKP) gauge bosons
$B^1 \to G^1+ \gamma$ and fermions $\psi^1\to G^1+\psi$, of relevance
to superWIMP dark matter. We then estimate the primordial abundance
for the tower of KK gravitons produced after reheating in
\secref{sec:reheating} and derive constraints on the reheat
temperature in \secref{sec:cosmology}.  Our conclusions are given in
\secref{sec:summary}.

\section{Graviton Interactions for Torus Compactifications}
\label{sec:torus}

To discuss graviton cosmology, we must first derive the interactions
of KK gravitons in UED.  Graviton interactions in extra dimensions
have been discussed previously, particularly in the context of
theories where SM fields are confined to four spacetime
dimensions. (See, \eg, Refs.~\cite{Han:1998sg,Giudice:1998ck}; we will
follow the analysis of Ref.~\cite{Han:1998sg} in discussing the
linearized gravitational action.)  Here we instead couple the
gravitational sector to extra-dimensional SM fields and also implement
the orbifold projection.

We begin in this section by determining the couplings of KK gravitons
and KK SM fields in the case of torus compactifications.  For $D$
spacetime dimensions, the action for matter coupled to gravity is
\begin{equation}
S^D = M_D^{D-2} \int d^D x \sqrt{|g|} \, \left[ R + {\cal L}
(\bar{\phi}) \right] \ ,
\end{equation}
where $M_D$ is the $D$-dimensional Planck scale and ${\cal L}$ is the
Lagrangian for all matter fields, which we denote generically by
$\bar{\phi}$. We use coordinates $x^M = (x^{\mu},y^i)$, where $0 \le
y^i \le 2 \pi R_i \equiv L_i$.

Linearizing about flat space using $g_{MN} = \eta_{MN} +
\bar{h}_{MN}$, where $\eta =(1, -1, -1, \ldots, -1)$,
we find the Fierz-Pauli action~\cite{Fierz:1939ix}
\begin{equation}
S^D_{\text{lin}} = M_D^{D-2} \int d^4 x \left(\prod_i\int_0^{L_i}
dy^i \right)\left[ K(\bar{h}) + {\cal L}_0 (\bar{\phi}) +
\frac{1}{2} \bar{h}_{MN} T^{MN} (\bar{\phi}) \right] \ ,
\end{equation}
where the graviton kinetic terms are contained in
\begin{equation}
K(\bar{h}) \equiv \frac{1}{4} \left( \bar{h}^{MN, P} \bar{h}_{MN,
P} - \bar{h}^{, M} \bar{h}_{, M} - 2 \bar{h}^{MN}{}_{, M}
\bar{h}_{PN}{}^{, P} + 2 \bar{h}_{, M} \bar{h}^{NM}{}_{, N}
\right) \ ,
\end{equation}
with $\bar{h} \equiv \bar{h}^M_{~M}$ and $\bar{h}^{MN} \equiv
\eta^{MP}\eta^{NQ}\bar{h}_{PQ}$. Also
\begin{equation}
{\cal L}_0 (\bar{\phi}) \equiv {\cal L} (\bar{\phi}) |_{g = \eta}
\end{equation}
is the flat-space matter Lagrangian, and
\begin{equation}
T^{MN} (\bar{\phi}) \equiv \left. \left( \eta^{MN} {\cal L} - 2
\frac{\partial {\cal L}} {\partial g_{MN}} \right) (\bar{\phi})
\right|_{g = \eta}
\end{equation}
is the flat-space stress-energy tensor.  To reproduce the standard
gravitational interactions among massless modes, we require
$M_D^{D-2}\prod_i L_i = (16 \pi G_N )^{-1} \equiv M_4^2$. Defining $h
= M_4 \bar{h}$ and $\phi = M_4 \bar{\phi}$, we find the linearized
action
\begin{eqnarray}
S^D_{\text{lin}} &=& \int d^4 x \left(\prod_i\int_0^{L_i}
\frac{dy^i}{L_i} \right)\left[ K(h) + {\cal L}_0 (\phi) +
\frac{1}{2 M_4} h_{MN} T^{MN} (\phi) \right] \ , \label{S5lin}
\end{eqnarray}
where the kinetic terms are now canonically normalized.

To write this in terms of 4D fields, we define
\begin{eqnarray}
h_{MN} &=& \left( \begin{array}{cc}
h_{\mu \nu} + \eta_{\mu \nu}\phi  & A_{\mu j} \\
A_{i \nu} & 2 \phi_{ij} \end{array} \right)\ , \\
T_{MN} &=& \left( \begin{array}{cc}
T_{\mu \nu} & T_{\mu j} \\
T_{i \nu}   & T_{ij} \end{array} \right) \ ,
\end{eqnarray}
where $\phi=\delta^{kl} \phi_{kl}$.  For simplicity, we will
henceforth take $R_i=R$ for all $i$.  Each component field is then
conventionally decomposed into 4D fields through the Fourier expansion
\begin{eqnarray}
h_{\mu \nu} (x, y) &=& \sum_{\vec{n}} h_{\mu \nu}^{\vec{n}} (x)
e^{i\vec{n} \cdot \vec{y}/R}\ , \nonumber \\
A_{\mu i} (x, y) &=& \sum_{\vec{n}} A_{\mu i}^{\vec{n}} (x)
e^{i\vec{n} \cdot \vec{y}/R}\ , \label{modes} \\
\phi_{ij} (x, y) &=& \sum_{\vec{n}} \phi_{ij}^{\vec{n}} (x)
e^{i\vec{n} \cdot \vec{y}/R}\ , \nonumber
\end{eqnarray}
and similarly for the stress-energy tensor component fields.  Here
$\vec{n} \cdot \vec{y} \equiv \sum_i n_i y^i$, where $\vec{n} = (n_1,
\ldots, n_d)$ and each $n_i$ runs from $-\infty$ to $\infty$. We
will also use the variable
\begin{equation}
m_n \equiv \left[ \frac{\delta^{ij} n_i n_j}{R^2} \right]^{1/2} \ .
\end{equation}

At each massive level ${\vec{n}}$, part of the fields $A_{\mu
j}^{\vec{n}}$, $A_{i \nu}^{\vec{n}}$, and $\phi_{ij}^{\vec{n}}$ are
absorbed to give mass to the graviton state $h_{\mu\nu}^{\vec{n}}$ in
an extra-dimensional realization of spontaneous symmetry breaking. The
resulting physical states are (defining ${\vec{n}}^2 \equiv
\delta^{kl} n_k n_l$)
\begin{eqnarray}
\tilde{h}_{\mu \nu}^{\vec{n}} &=& h_{\mu \nu}^{\vec{n}} + i
{\delta^{ij} n_i R\over {\vec{n}}^2} \left(
\partial_{\mu} A_{\nu j}^{\vec{n}} +
\partial_{\nu} A_{\mu j}^{\vec{n}} \right)
- \left(\delta_{ij}+2{n_in_j \over {\vec{n}}^2} \right)\left(
{2\over 3} \frac{\partial_{\mu}
\partial_{\nu} }{m_n^2}
- {\eta_{\mu \nu}\over 3} \right) \phi_{ij}^{\vec{n}} \ , \\
\tilde{A}_{\mu i}^{\vec{n}} &=& P_{ij}\left( {A}_{\mu j}^{\vec{n}}
+ 2i {\delta^{kl} n_k R\over {\vec{n}}^2}
\partial_\mu\phi^{\vec{n}}_{jl} \right) \ , \\
\tilde{\phi}_{ij}^{\vec{n}}
&=& \sqrt{2}(P_{ik}P_{jl}+aP_{ij}P_{kl})\phi^{\vec{n}}_{kl}\ ,
\end{eqnarray}
where we have defined
\begin{eqnarray}
P_{ij}=\left(\delta_{ij}-{n_in_j\over {\vec{n}}^2} \right)\ ,
\end{eqnarray}
and $a$ satisfies $3(d-1)a^2+6a=1$.

The Lagrangian for the physical fields needs to be treated slightly
differently for the zero modes versus the non-zero modes.  For the
zero modes, the action is
\begin{equation}
S^0_{\text{lin}} = \int d^4 x \left[ {\cal L}_h^0 + {\cal L}_{h\,
\text{int}}^0 \right] \ ,
\end{equation}
where
\begin{eqnarray}
{\cal L}_h^0 &=& \frac{1}{4} \left( h^{0\, \mu \nu , \rho} h^0_{\mu
\nu, \rho} - h^0_{, \mu} h^{0\, , \mu} - 2 h^{0\, \mu \nu}{}_{,
\mu} h^0_{\rho \nu}{}^{, \rho} + 2 h^0_{, \mu} h^{0\, \nu \mu}{}_{,
\nu} \right)  \nonumber \\
&& - \frac{1}{4}\sum_i F^0_{\mu\nu\, i} F^{0\, \mu\nu\, i} +
\frac{1}{2} \partial^{\mu} \phi^0 \partial_{\mu} \phi^0 +
\sum_{ij}
\partial^{\mu} \phi^0 _{ij} \partial_{\mu} \phi^0 _{ij} \ ,\\
{\cal L}^0_{h\, \text{int}} &=& \frac{1}{2 M_4} h^0_{\mu \nu} T^{0\, \mu
\nu} +\frac{1}{ M_4} A^0_{\mu i} T^{0\, \mu i}+\frac{1}{2 M_4}
\phi^0_{ij} T^{0\,ij} \ ,
\end{eqnarray}
and $F^0_{\mu \nu\, i} \equiv \partial_{\mu} A^0_{\nu i} -
\partial_{\nu} A^0_{\mu i}$.

The non-zero modes have the kinetic terms
\begin{equation}
S^{\vec{n}\, \text{K.E.}}_{\text{lin}}
= \int d^4 x  \sum_{\vec{n} \ne \vec{0} }
\left[ {\cal L}_h^{\vec{n}} + {\cal L}_A^{\vec{n}}
+ {\cal L}_\phi^{\vec{n}} \right] \ ,
\label{ssum}
\end{equation}
where
\begin{eqnarray}
{\cal L}_h^{\vec{n}} &=& \frac{1}{4} \left( \tilde{h}^{{\vec{n}}\,
\mu \nu , \rho} \tilde{h}^{-{\vec{n}}}_{\mu \nu, \rho} -
\tilde{h}^n_{, \mu} \tilde{h}^{-{\vec{n}}\, , \mu} - 2
\tilde{h}^{{\vec{n}}\, \mu \nu}{}_{, \mu}
\tilde{h}^{-{\vec{n}}}_{\rho \nu}{}^{, \rho}
+ 2 \tilde{h}^{\vec{n}}_{, \mu}
\tilde{h}^{-{\vec{n}}\, \nu \mu}{}_{, \nu} \right)
\nonumber \\
&& - \frac{1}{4} m_n^2 \tilde{h}^{{\vec{n}}\, \mu \nu}
\tilde{h}^{-{\vec{n}}}_{\mu \nu}
+ \frac{1}{4} m_n^2 \tilde{h}^{\vec{n}} \tilde{h}^{-{\vec{n}}} \ , \\
{\cal L}_A^{\vec{n}} &=& -{1\over 4}
\tilde{F}_{\mu\nu\, i}^{\vec{n}} \tilde{F}^{-{\vec{n}}\, \mu\nu\, i }
+ \frac{1}{2} m_n^2\tilde{A}_{\mu i}^{\vec{n}}
\tilde{A}^{-{\vec{n}}\, \mu i}\ , \\
{\cal L}_\phi^{\vec{n}} &=& {1\over
2}\partial_\mu\tilde{\phi}^{\vec{n}}_{ij}
\partial^\mu\tilde{\phi}^{-{\vec{n}}\, ij}
- \frac{1}{2} m_n^2
\tilde{\phi}^{\vec{n}}_{ij}\tilde{\phi}^{-{\vec{n}}\, ij}\ ,
\end{eqnarray}
and $\tilde{F}^{\vec{n}}_{\mu \nu\, i} \equiv \partial_{\mu}
\tilde{A}^{\vec{n}}_{\nu i} - \partial_{\nu} \tilde{A}^{\vec{n}}_{\mu
i}$.  The interaction terms of the non-zero modes are
\begin{equation}
S^{\vec{n}\, \text{int}}_{\text{lin}}
= \int d^4 x \sum_{\vec{n} \ne \vec{0}}
\left[ {\cal L}_{h\, \text{int}}^{\vec{n}}
+ {\cal L}_{A\, \text{int}}^{\vec{n}}
+ {\cal L}_{\phi\, \text{int}}^{\vec{n}} \right] \ ,
\label{ssum2}
\end{equation}
where
\begin{eqnarray}
{\cal L}^{\vec{n}}_{h\, \text{int}} &=& \frac{1}{2M_4}
\tilde{h}^{\vec{n}}_{\mu \nu} T^{-{\vec{n}}\, \mu \nu} \ , \\
{\cal L}^{\vec{n}}_{A\, \text{int}} &=& \frac{1}{M_4} P_{ij}
\tilde{A}^{\vec{n}\, \mu j}T^{-{\vec{n}}\, i}_{\mu} \ , \\
{\cal L}^{\vec{n}}_{\phi\, \text{int}} &=& \frac{1}{2M_4}
\tilde{\phi}^{\vec{n}}_{ij}\tilde T^{-{\vec{n}}\, ij} \ ,
\end{eqnarray}
with
\begin{eqnarray}
\tilde T^{\vec{n}}_{ij}
= \left[ {2\over 3}\eta^{\mu\nu} T^{\vec{n}}_{\mu\nu}
+ {2\over 3\partial_k^2} \partial^\nu
( \partial^\mu T^{\vec{n}}_{\mu\nu} )
+ 2 a P_{kl}T^{\vec{n}}_{kl} \right]
{P_{ij} \over\sqrt{2}(1+a(d-1))}
- \sqrt{2}P_{ik}P_{jl}T^{\vec{n}}_{kl}\ .
\end{eqnarray}
Note that in all expressions above, both $(n_1, n_2, \ldots, n_d)$ and
$(-n_1, -n_2, \ldots, -n_d)$ are to be included in sums over $\vec{n}
\ne \vec{0}$, and all $d^2$ pairs are to be included in sums over
$ij$.

\section{Graviton Interactions for Orbifold Compactifications}
\label{sec:orbifold}

The analysis so far is valid for compactification on a torus $S^d$.
For reasons described in \secref{sec:introduction}, we are most
interested in the case where KK number is broken to KK-parity.  This
may be achieved by compactifications on orbifolds.

For simplicity, we now concentrate on the $D=5$ case with coordinates
$x^M = (x^{\mu},y)$, where the extra dimension is an interval
$S^1/Z_2$ of length $\pi R$~\cite{Appelquist:2000nn}.  This geometry
can be described as an orbifold of a circle compactification, where we
orbifold by the symmetry $y \to -y$.  In the SM sector, this symmetry
is accompanied by the transformations $V_\mu \to V_\mu$ and $V_5 \to
-V_5$ for the gauge fields. The 4D scalar is projected out, and the
massless sector includes only the 4D gauge field.  A similar
projection on fermions removes half of the degrees of freedom, leaving
only the chiral fermions of the SM.

In the gravitational sector, we will project by the action $h_{\mu\nu}
\to h_{\mu\nu}$, $h_{\mu 5} \to - h_{\mu 5}$, and $h_{55} \to h_{55}$
under $y \to -y$.  At the massless level, this preserves the 4D
graviton $h_{\mu\nu}^0(x)$, while removing the gravi-vectors $h_{\mu
5}^0(x)$ and $h_{5 \nu}^0(x)$. The gravi-scalar $h_{55}^0(x)$ is
preserved by this projection; we assume that some other physics
stabilizes this mode and generates a large mass for it.  (The
phenomenology and cosmology of an extremely light gravi-scalar is
discussed in Refs.~\cite{Perivolaropoulos:2002pn}.)  The final
massless sector is, then, just the SM plus the 4D graviton.

\subsection{Boundary Interactions}
\label{sec:boundary}

Prior to discussing the bulk interactions, we need to consider the
boundary terms. These arise because the space we are considering is
singular at $y=0,\pi R$, and hence we expect new interactions at these
points. The nature of these interactions can only be derived from an
underlying microscopic theory (such as string theory) that smooths out
the singularity.

Nevertheless, we can make some qualitative statements about these
terms. If the singular points are smoothed out to a size $l_F$, these
interactions are localized near the singular points in a region of
size $l_F$. We can formally write these interactions in the form
\begin{eqnarray}
S=\int d^4x\int_0^{\pi R}{dy\over \pi R}
\left[ f\left( {y\over l_F} \right)+
f\left( {\pi R-y\over l_F}\right)\right] {\cal L}(x)\ ,
\end{eqnarray}
where $f(w)$ goes to zero for $w \gg 1$. We have assumed that the
resolution of the singularity is such that KK-parity is preserved.

The basic assumption we will make is that $l_F \ll \pi R$. This is not
unreasonable; for example, a natural guess for the scale $l_F$ would
be the (higher-dimensional) Planck scale. If we then consider
$R^{-1}\sim 1~\tev$, we find $l_F^{-1}\sim 10^{10}~\tev$, so indeed
$l_F \ll \pi R$ in this case.  If $l_F \ll \pi R$, we can estimate the
boundary terms as
\begin{eqnarray}
S\sim {l_F\over \pi R} \, f(0)\int d^4 x \, {\cal L}(x) \ ,
\end{eqnarray}
that is, the boundary terms are suppressed by a factor $l_F/\pi R$,
which is small. Thus, to leading order, we can ignore these terms.

KK number-violating terms are generated at
one-loop~\cite{Georgi:2000ks,Cheng:2002iz}. The tree-level spectrum
already breaks KK number, since certain modes are projected out by the
orbifold action.  These are translated at loop-level to interactions
that violate KK number. The typical decay widths of the higher KK
modes are therefore $\sim \alpha m_n$.

There is another source of boundary terms: certain manipulations
of the bulk terms require integration by parts, which produce
boundary contributions. These terms are also suppressed for the
reason given above.

\subsection{Bulk Interactions}
\label{sec:bulk}

Because we project out modes that are even or odd under $y \to -y$, it
is more convenient to replace the conventional Fourier expansion of
\eqref{modes} by the expansion
\begin{eqnarray}
h_{\mu \nu}(x, y) &=& h_{+\, \mu\nu}^0(x) 
+ \sqrt{2} \sum_{n>0} \left[ h_{+\, \mu\nu}^n(x) \cos \frac{ny}{R} 
+ h_{-\, \mu\nu}^n(x) \sin \frac{ny}{R} \right] \ ,\\
A_{5 \nu}(x, y) &=& A_{+\, 5\nu}^0 (x) 
+ \sqrt{2} \sum_{n>0} \left[ A_{+\, 5\nu}^n(x) \cos \frac{ny}{R} 
+ A_{-\, 5\nu}^n(x) \sin \frac{ny}{R} \right] \ ,\\
\phi_{5 5}(x, y) &=& \phi_{+\, 55}^0 (x) 
+ \sqrt{2} \sum_{n>0} \left[ \phi_{+\, 55}^n(x) \cos \frac{ny}{R} 
+ \phi_{-\, 55}^n(x) \sin \frac{ny}{R} \right] \ ,\\
T_{\mu \nu}(x, y) &=& T^0_{\mu \nu}(x) + \sqrt{2} \sum_{n>0} \left[
T^n_{+\, \mu \nu}(x) \cos \frac{ny}{R} + T^n_{-\, \mu \nu}(x) \sin
\frac{ny}{R} \right] \ ,
\end{eqnarray}
and similarly for the other component fields. All $h_{-\, \mu\nu}^n$,
$A_{+\, 5\nu}^n$, $A_{+\, \mu 5}^n$, and $\phi_{-\, 55}^n$ fields are
projected out by the orbifold.

The physical graviton field $G^n$ at KK level $n>0$ is
\begin{eqnarray}
G^n_{\mu\nu} = 
\left( h_{+\, \mu\nu}^n + \eta_{\mu\nu} \phi_{+\, 55}^n \right)
+ \frac{R}{n} \left( \partial_{\mu} A_{-\, 5\nu}^n
+ \partial_{\nu} A_{-\, 5 \mu}^n \right)
- \frac{R^2}{n^2} \partial_{\mu} \partial_{\nu} 
\left( 2 \phi_{+\, 55}^n \right) \ .
\end{eqnarray}

In this new basis, the linearized action for the $G$ fields is
\begin{eqnarray}
S^5_{\text{lin}} &=& \int d^4x \Biggl[ {\cal L}^0_G + {\cal L}_0
(\bar{\phi}) + {\cal L}_{G\, \text{int}}^0
+ \sum_{n > 0} \left( {\cal L}_G^n + {\cal L}_{G\,
\text{int}}^n
\right) \Biggr] \ ,
\end{eqnarray}
where
\begin{eqnarray}
{\cal L}_G^0 &=& \frac{1}{4} \left( G^{0\, \mu \nu , \rho}
G^0_{\mu \nu, \rho} - G^0_{, \mu} G^{0\, , \mu} - 2 G^{0\, \mu
\nu}{}_{, \mu} G^0_{\rho \nu}{}^{, \rho}
+ 2 G^0_{, \mu} G^{0\, \nu \mu}{}_{, \nu} \right)\ , \\
{\cal L}^0_{G\, \text{int}} &=& \frac{1}{2 M_4} G^0_{\mu \nu}
T^{0\, \mu \nu}\ , \\
{\cal L}_G^n &=& \frac{1}{4} \left( G^{n\, \mu \nu , \rho}
G^n_{\mu \nu, \rho} - G^n_{, \mu} G^{n\, , \mu} - 2 G^{n\, \mu
\nu}{}_{, \mu} G^n_{\rho \nu}{}^{, \rho} + 2 G^n_{, \mu} G^{n\,
\nu \mu}{}_{, \nu} \right) \ ,
\nonumber \\
&& - \frac{1}{4} m_n^2 G^{n\, \mu \nu}
G^{n}_{\mu \nu} + \frac{1}{4} m_n^2 G^n G^n \ ,\\
{\cal L}^n_{G\, \text{int}} &=& \frac{1}{2 M_4} G^n_{\mu \nu}
T_+^{n\, \mu \nu}\ .
\end{eqnarray}

The KK graviton propagator is\footnote{This analysis differs by a
factor of 2 from the journal version of Ref.~\cite{Han:1998sg}, but
agrees with the later e-print version hep-ph/9811350v4.}
\begin{equation}
\langle G^m_{\mu\nu} G^n_{\rho\sigma} \rangle = \frac{i
\delta^{mn} B_{\mu\nu\, \rho\sigma} (k)} {k^2 - m_n^2 + i
\varepsilon} \ ,
\end{equation}
where
\begin{eqnarray}
B_{\mu\nu\, \rho\sigma}(k) &=& 2\left(\eta_{\mu\rho}-\frac{k_\mu
k_\rho}{m_n^2}\right) \left(\eta_{\nu\sigma}-\frac{k_\nu
k_\sigma}{m_n^2}\right) +2\left(\eta_{\mu\sigma}-\frac{k_\mu
k_\sigma}{m_n^2}\right)
\left(\eta_{\nu\rho}-{k_\nu k_\rho\over m_n^2}\right)\nonumber\\
&& - \frac{4}{3} \left(\eta_{\mu\nu}-\frac{k_\mu
k_\nu}{m_n^2}\right) \left(\eta_{\rho\sigma}-\frac{k_\rho
k_\sigma}{m_n^2}\right)\ ,
\label{B}
\end{eqnarray}
and $m_n=n/R$.

To determine the KK graviton couplings, we must determine the matter
stress-energy tensor components $T_+^{n\, \mu \nu}$. For gauge fields,
\begin{equation}
T_{MN} = F_M{}^P F_{PN} - \frac{1}{4} \eta_{MN} F_{PQ} F^{PQ}
\ .
\end{equation}
We expand the gauge field in harmonics
\begin{eqnarray}
a_{\mu}(x,y) &=& a_{+\, \mu}^0(x) 
+ \sqrt{2} \sum_{n>0} \left[ a_{+\, \mu}^n (x) \cos \frac{ny}{R}
+a_{-\, \mu}^n (x) \sin \frac{ny}{R} \right] \ , \\
a_5(x,y) &=& a_{+\, 5}^0(x)
+ \sqrt{2} \sum_{n>0} \left[ a_{+\, 5}^n (x) \cos \frac{ny}{R}
+ a_{-\, 5}^n (x) \sin \frac{ny}{R}\right] \ .
\end{eqnarray}
All $a_{-\, \mu}^n$ and $a_{+\, 5}^n$ fields are projected out by the
orbifold.

The physical gauge field $A^n_{\mu}$ at KK level $n>0$ is
\begin{equation}
A^n_{\mu}= a_{+\, \mu}^n - \frac{R}{n} \partial_{\mu} a_{-\, 5}^n \ ,
\end{equation}
and identifying coefficients of $\cos ny/R$, we find
\begin{equation}
T_{+\, \mu \nu}^n = \sum_{m=0}^n\left[ F_{\mu}{}^{m\, \rho}
F_{\nu\rho}^{n-m} - \frac{1}{4} \eta_{\mu\nu} F^m_{\rho\sigma}
F^{n-m\, \rho\sigma} +m_nm_{n-m}(A_{\mu}^mA_{\nu}^{n-m}
-\frac{1}{2}\eta_{\mu\nu}A_{\rho}^nA^{n-m\, \rho})\right] ,
\end{equation}
where $F^m_{\mu\nu} \equiv \partial_{\mu} A^m_{\nu} - \partial_{\nu}
A^m_{\mu}$. The $G^n_{\mu\nu}(q) A^m_{\alpha}(k_1)
A^{n-m}_{\beta}(k_2)$ vertex is therefore
\begin{eqnarray}
X_{\mu\nu\alpha\beta} &=& \frac{i}{2M_4}\biggl[\eta_{\alpha\beta}
k_{1 \mu} k_{2 \nu}
-\eta_{\mu\alpha}k_{1\beta}k_{2\nu}-\eta_{\nu\beta}k_{1\mu}k_{2\alpha}
+\eta_{\mu\alpha}\eta_{\nu\beta}(k_1\cdot k_2) \nonumber \\
&& -\frac{1}{2}\eta_{\mu\nu}\left( \eta_{\alpha\beta}(k_1\cdot k_2)
 -k_{1\beta}k_{2\alpha} \right) +
m_n m_{n-m} ( \eta_{\mu\alpha}\eta_{\nu\beta}
-\frac{1}{2} \eta_{\mu\nu}\eta_{\alpha\beta} ) + \left(\alpha
\leftrightarrow \beta \right) \biggr] .
\end{eqnarray}

The fermion couplings may be calculated similarly.  The stress-energy
tensor for 5D Dirac fermions is
\begin{eqnarray}
T_{MN} &=& \eta_{MN}(\overline{\psi}
i\gamma^{P}D_P\psi-m_{\psi^0}\overline{\psi}\psi)
-\frac{1}{2}\overline{\psi}i\gamma_MD_N\psi-
\frac{1}{2}\overline{\psi}i\gamma_ND_M\psi\nonumber\\
&&-\frac{1}{2}\eta_{MN}\partial^P(\overline{\psi}i\gamma_P\psi)
+\frac{1}{4}\partial_M(\overline{\psi}i\gamma_N\psi)
+\frac{1}{4}\partial_N(\overline{\psi}i\gamma_M\psi)\ ,
\end{eqnarray}
where $m_{\psi^0}$ is the 5-dimensional mass of the fermion.  We
will consider a $Z_2$ action on the fermion which acts as
$\psi(x,y)=-\gamma_5\psi(x,-y)$. This preserves the zero mode of
the left handed fermion, but not that of the right handed one. In
this case, the expansion in harmonics is of the form
\begin{eqnarray}
\psi(x,y) &=& \psi^0_L(x)+\sqrt{2} \sum_{n>0} \left[ \psi^n_L(x)
\cos\frac{ny}{R}+i\psi^n_R(x)\sin\frac{ny}{R} \right] \ , \\
i\partial_5\gamma_5\psi(x,y) &=&
\sqrt{2}\sum_{n>0} im_n \left[ \psi^n_L(x)
\sin\frac{ny}{R}+i\psi^n_R(x)\cos\frac{ny}{R} \right] \ .
\end{eqnarray}
The stress-energy tensor for fermions is
\begin{eqnarray}
T_{+\, \mu\nu}^n &=& \sum_{m=0}^n \biggl[
\eta_{\mu\nu}(\overline{\psi_L^m}
i\gamma^{\rho}D_{\rho}\psi_L^{n-m} -m_{n-m}
\overline{\psi_R^m}\psi_L^{n-m})
\nonumber\\
&&-\frac{1}{2}\overline{\psi^m_L}i\gamma_{\mu}D_{\nu}\psi^{n-m}_L
-\frac{1}{2}\overline{\psi^m_L}i\gamma_{\nu}D_{\mu}\psi_L^{n-m}\nonumber\\
&&-\frac{1}{2}\eta_{\mu\nu}\partial^{\rho}
(\overline{\psi^m_L}i\gamma_{\rho}\psi_L^{n-m})
+\frac{1}{2}\eta_{\mu\nu}(m_m+m_{n-m})
(\overline{\psi^m_R}\psi_L^{n-m})\nonumber\\
&&+\frac{1}{4}\partial_{\mu}
(\overline{\psi_L^m}i\gamma_{\nu}\psi_L^{n-m})
+\frac{1}{4}\partial_{\nu}
(\overline{\psi_L^m}i\gamma_{\mu}\psi_L^{n-m})+(R\leftrightarrow L)
\biggr] ,
\end{eqnarray}
where $\psi^0_R(x)=0$.  The KK graviton-fermion-fermion vertex for
$G^n_{\mu\nu}(q) \bar{\psi}^m(k_1) \psi^{n-m}(k_2)$ interactions is
then
\begin{eqnarray}
Y_{\mu\nu}&=& \frac{i}{4M_4}\biggl[ \eta_{\mu\nu}
\left[ 2(\gamma^{\rho}k_{2\rho}-m_{n-m})-(\gamma^{\rho}k_{1\rho}-m_m)
-(\gamma^{\rho}k_{2\rho}-m_{n-m}) \right] \nonumber\\
&&  \quad -\frac{1}{2}(k_1+k_2)_{\mu}\gamma_{\nu}-
\frac{1}{2}(k_1+k_2)_{\nu}\gamma_{\mu}\biggr] \ .
\end{eqnarray}
Note that if the fermions in this vertex are on-shell, only the
last two terms remain.

\section{NLKP Decay Widths into LKP Gravitons}
\label{sec:widths}

Using the vertices of \secref{sec:bulk}, we can now calculate Feynman
diagrams involving gravitons and SM fields. We are particularly
interested in the decay widths for applications to models of dark
matter~\cite{Feng:2003xh,Feng:2003uy}, where an NLKP SM field decays
to a stable LKP graviton. The NLKP may either be a gauge boson, such
as $B^1$, or a fermion, such as $\tau^1$.  These decay widths may be
calculated using the usual trace techniques or by helicity amplitude
methods.  We have done both and checked that they yield identical
answers.  We present the helicity amplitude analysis here.

The polarization vectors of a massive graviton, given the
normalization conventions of \secref{sec:bulk}, are
\begin{eqnarray}
e_{\mu\nu}^{\pm 2}&=&2\epsilon_{\mu}^{\pm}\epsilon_{\nu}^{\pm}
\ , \nonumber\\
e_{\mu\nu}^{\pm 1}&=&\sqrt{2}\, (\epsilon_{\mu}^{\pm}\epsilon_{\nu}^{0}+
\epsilon_{\mu}^{0}\epsilon_{\nu}^{\pm})\ ,\nonumber\\
e_{\mu\nu}^{0}&=&\sqrt{\frac{2}{3}}\, (\epsilon_{\mu}^{+}
\epsilon_{\nu}^{-}+
\epsilon_{\mu}^{-}\epsilon_{\nu}^{+}-
2\epsilon_{\mu}^{0}\epsilon_{\nu}^{0})\ . \nonumber
\end{eqnarray}
Here $\epsilon_{\mu}^{\pm}$ and $\epsilon_{\mu}^{0}$ are the
polarization vectors of a massive gauge boson; for a massive vector
boson with $p^{\mu}=(E,0,0,p)$ and mass $m$,
\begin{eqnarray}
 \epsilon^+_{\mu}(p)&=&\frac{1}{\sqrt{2}}(0,1,i,0)\ ,\\
 \epsilon^-_{\mu}(p)&=&\frac{1}{\sqrt{2}}(0,-1,i,0)\ ,\\
 \epsilon^0_{\mu}(p)&=&\frac{1}{m}(p,0,0,-E)\ .
\end{eqnarray}
The graviton polarization vectors satisfy the normalization and
polarization sum conditions
\begin{eqnarray}
e^{s\,\mu\nu}e^{s'\, *}_{\mu\nu} &=& 4\delta^{s s'}\ ,\\
\sum_s e^s_{\mu\nu}e^{s\, *}_{\rho\sigma} &=& B_{\mu\nu\, \rho\sigma}\ ,
\end{eqnarray}
where $B_{\mu\nu\, \rho \sigma}$ is as given in \eqref{B}.

With these conventions, the helicity amplitudes for the decay of a KK
hypercharge gauge boson to a KK graviton are
\begin{eqnarray}
\left| {\cal M} \left(
B^{1\, \pm}(k) \to G^{1\,0}(p)+\gamma^{\pm}(q)
\right) \right|
&=& \cos \theta_W X^{\mu\nu\alpha\beta} \epsilon_{\alpha}^{\pm}(k)
\epsilon_{\beta}^{\pm\, *}(q) e_{\mu\nu}^{0\, *}(p) \nonumber\\
&=&\frac{\cos \theta_W}{\sqrt{6} M_4} \frac{m_{B^1}^4}{m_{G^1}^2}
\left[ 1-\frac{m_{G^1}^2}{m_{B^1}^2} \right] , \\
\left| {\cal M} \left(
B^{1\,0}(k)\to G^{1\,\pm 1}(p)+\gamma^{\mp}(q)
\right) \right|
&=& \cos \theta_W X^{\mu\nu\alpha\beta}\epsilon_{\alpha}^{0}(k)
\epsilon_{\beta}^{\mp\, *}(q) e_{\mu\nu}^{\pm 1 \, *}(p)\nonumber\\
\qquad &=&\frac{\cos \theta_W}{\sqrt{2}M_4} \frac{m_{B^1}^3}{m_{G^1}}
\left[ 1-\frac{m_{G^1}^2}{m_{B^1}^2} \right] , \\
\left| {\cal M} \left(
B^{1\,\pm}(k) \to G^{1\, \pm 2}(p)+\gamma^{\mp}(q)
\right) \right|
&=& \cos \theta_W X^{\mu\nu\alpha\beta}\epsilon_{\alpha}^{\pm}(k)
\epsilon_{\beta}^{\pm\, *}(q) e_{\mu\nu}^{\pm 2\, *}(p)
\nonumber\\
&=&\frac{\cos \theta_W}{M_4}m_{B^1}^2
\left[ 1-\frac{m_{G^1}^2}{m_{B^1}^2} \right] .
\end{eqnarray}
The total squared amplitude, averaged over the initial three
polarizations and summed over final states, is
\begin{eqnarray}
\lefteqn{
\frac{1}{3} \sum_{h, h', h''} \left| {\cal M} \left(
B^{1\, h} (k) \to G^{1\, h'}(p) + \gamma^{h''}(q) \right) \right|^2
} \nonumber \\
&& \quad = \frac{\cos^2\theta_W}{9}\frac{m_{B^1}^4}{M_4^2}
\left[ 1-\frac{m_{G^1}^2}{m_{B^1}^2} \right]^2
\left[ 6+3\frac{m_{B^1}^2}{m_{G^1}^2}+ \frac{m_{B^1}^4}{m_{G^1}^4}
\right] \ ,
\end{eqnarray}
and the decay width of a KK hypercharge gauge boson to a KK graviton
is therefore
\begin{equation}
\Gamma(B^1 \to G^1 \gamma) = \frac{\cos^2\theta_W}{144\pi M_4^2}
\frac{m_{B^1}^7}{m_{G^1}^4} \left[1 - \frac{m_{G^1}^2}{m_{B^1}^2}
\right]^3
\left[1 + 3 \frac{m_{G^1}^2}{m_{B^1}^2}
+ 6 \frac{m_{G^1}^4}{m_{B^1}^4} \right] \ .
\label{B1lifetime}
\end{equation}

The decay width of a chiral KK fermion to a KK graviton may be
calculated similarly in the helicity amplitude formalism.  For
$p^{\mu}=(E,0,0,p)$, the helicity spinors are
\begin{equation}
 u^+(p)=\left(\begin{array}{c}
   \sqrt{E+m}\ \xi^+\\
   \sqrt{E-m}\ \xi^+
 \end{array}\right)\ , \qquad
 u^-(p)=\left(\begin{array}{c}
   \sqrt{E+m}\ \xi^-\\
   -\sqrt{E-m}\ \xi^-
 \end{array}\right)\ ,
\end{equation}
where $\xi^{+\, T}=(1,0)$ and $\xi^{-\, T}=(0,1)$, and we take the
Dirac representation.

The helicity amplitudes for a chiral fermion decaying to a KK graviton
are, then,
\begin{eqnarray}
\left| {\cal M} \left(
f^{1\,\pm}_{L,R}(k)\to G^{1\,\pm 1}(p)+f^{0\,\mp}_{L,R}(q)
\right) \right|
&=& Y^{\mu\nu} \overline{u^{\mp}(q)} u^{\pm}(k) e_{\mu\nu}^{\pm 1\, *}(p)
\nonumber\\
&=&\frac{1}{2M_4}\frac{m_{f^1}^3}{m_{G^1}}
\left[ 1-\frac{m_{G^1}^2}{m_{f^1}^2} \right]^{3/2}
\ , \\
\left| {\cal M} \left(
f^{1\,\pm}_{L,R}(k) \to G^{1\,0}(p)+f^{0\,\pm}_{L,R}(q)
\right) \right|
&=&Y^{\mu\nu} \overline{u^{\pm}(q)} u^{\pm}(k) e_{\mu\nu}^0 (p)
\nonumber\\
&=&\frac{1}{\sqrt{6}M_4}\frac{m_{f^1}^4}{m_{G^1}^2}
\left[ 1-\frac{m_{G^1}^2}{m_{f^1}^2} \right]^{3/2}
\ ,
\end{eqnarray}
where we have assumed a massless zero mode fermion.  The squared
amplitude, averaged over the initial two polarizations and summed over
final states, is
\begin{eqnarray}
\lefteqn{
\frac{1}{2} \sum_{h, h', h''} \left| {\cal M} \left(
f^{1\, h}_{L,R}(k)\to G^{1\, h'}(p)+f^{0\, h''}_{L,R}(q)
\right) \right|^2 }
\nonumber \\
&& \quad = \frac{1}{6}\frac{m_{f^1}^4}{M_4^2}
\left[ 1-\frac{m_{G^1}^2}{m_{f^1}^2} \right]^3
\left[ 3+2\frac{m_{f^1}^2}{m_{G^1}^2} \right]
\frac{m_{f^1}^2}{m_{G^1}^2} \ .
\end{eqnarray}
The decay width of a chiral fermion to a KK graviton is
\begin{equation}
\Gamma(f_{L,R}^1 \to f_{L,R}^0+G^1) =
\frac{1}{96\pi M_4^2}\frac{m_{f^1}^7}{m_{G^1}^4}
\left[1 - \frac{m_{G^1}^2}{m_{f^1}^2}\right]^4
\left[2+3 \frac{m_{G^1}^2}{m_{f^1}^2}\right]\ .
\label{f1lifetime}
\end{equation}

The decay widths of \eqsref{B1lifetime}{f1lifetime} provide accurate
expressions for NLKP lifetimes in the LKP graviton scenario.  For NLKP
and LKP masses at the weak scale $\mweak$, and assuming mass
splittings of the same order, the naive expectation of lifetimes $\tau
= \Gamma^{-1} \sim 4 \pi M_4^2/(\mweak)^3 \sim 10^5~\s - 10^8~\s$ is
born out.  NLKPs therefore decay after BBN, and their decay products
are subject to rather stringent BBN constraints.

Even relatively mild degeneracies may disrupt these expectations,
however, as for $\Delta m \equiv m_{\text{NLKP}} - m_{\text{LKP}} \ll
m_{\text{LKP}}$ the lifetimes scale as $\tau \propto (\Delta m)^{-3}$
and $\tau \propto (\Delta m)^{-4}$ in the gauge boson and fermion
cases, respectively. This behavior is mirrored in the analogous case
in supersymmetry of superpartners decaying to gravitinos.  The
supersymmetric case was analyzed in Ref.~\cite{Feng:2003uy}, where
alternative signals of superWIMP dark matter were identified.  The
similarity of \eqsref{B1lifetime}{f1lifetime} to the supersymmetric
case indicates that KK graviton dark matter is also a viable superWIMP
candidate with promising possibilities for detection.

\section{Reheating and KK Graviton Production}
\label{sec:reheating}

KK gravitons are produced copiously after the Big Bang. Inflation
dilutes these gravitons away, but their number density may be
regenerated during reheating.  The situation is analogous to the case
of supersymmetry, where gravitinos may be produced after reheating,
and constraints bound the reheat temperature to $\TRH \alt 10^8 -
10^{10}~\gev$~\cite{Pagels:ke,Bolz:1998ek}. As we will see, however,
the presence of a tower of KK levels leads to much stronger production
of gravitons, and much stronger bounds on reheat temperature in UED
models.  In this section we estimate the density of KK gravitons
produced after reheating for UED models with an arbitrary number of
extra dimensions $d$.  In the next section, we discuss the
cosmological significance of these results.  Throughout we assume that
the extra dimensions remain fixed in size.

UED models are characterized by two hierarchically separated scales:
the KK mass scale $\mKK \sim \tev$, and the 4D Planck scale $M_4
\simeq 1.7\times 10^{18}~\gev$.  For this analysis, cosmology
introduces another scale, the reheat temperature $\TRH$.  As UED
models are 4D effective theories of some higher-dimensional theory,
they are valid only up to some cutoff scale $M_s$, which is much
smaller than $M_4$ \cite{Chivukula:2003kq}.  We therefore assume $\TRH
< M_s$, and so $\TRH \ll M_4$.

In terms of these scales, the expansion rate of the universe is $H
\sim T^2/M_4$.  The interaction rate of SM particles and their KK
partners with each other is $\sigma_{\text{SM}} n \sim \alpha^2 T$.
The decay rate of SM particles at KK level $n$ may also be estimated
to be $\Gamma_{\text{SM}} \sim \alpha m_n$.  Given the hierarchy
between $T$ and $M_4$,
\begin{equation}
\sigma_{\text{SM}} n \, , \Gamma_{\text{SM}} \gg H \ ,
\end{equation}
and so SM particles and their KK partners remain in thermodynamic
equilibrium as the universe cools after reheating.  In contrast, the
time scale for graviton interactions is very long. For example, the
interaction rate for $\phi_a^{\vec{k}} + \phi_b^{\vec{l}} \to
\phi_c^{\vec{m}} + G^{\vec{n}}$, where $\vec{k}$, $\vec{l}$,
$\vec{m}$, and $\vec{n}$ specify KK levels and $a$, $b$, and $c$ label
SM degrees of freedom, is $\sigma_G n \sim T^3/M_4^2$.  Similarly, as
discussed in \secref{sec:widths}, the decay rates of KK gravitons are
$\Gamma_G \sim m_n^3/M_4^2$.  These rates are therefore far below the
expansion rate,
\begin{equation}
\sigma_G n \, , \Gamma_G \ll H \ ,
\end{equation}
and gravitons never reach thermodynamic equilibrium.

The qualitative picture, then, is that after reheating, the SM degrees
of freedom exist in thermal equilibrium.  Occasionally, they produce
KK gravitons, which are meta-stable and decay long after Big Bang
nucleosynthesis.  If overproduced, they may overclose the universe or
their eventual decay products may destroy the predictions of Big Bang
nucleosynthesis.

We now work toward a more quantitative estimate of graviton
abundances.  During the era of graviton production, the expansion rate
$H$ is given by
\begin{equation}
H^2 = \frac{8 \pi G_N}{3} \rho_R = \frac{\pi^2}{180 M_4^2} g_*(T) T^4
\ ,
\label{H}
\end{equation}
where $\rho_R$ and $g_*(T)$ are the total energy density and the
effective number of light degrees of freedom, respectively.  The
entropy density is
\begin{equation}
s = \frac{2 \pi^2}{45} g_*(T) T^3 \ ,
\label{s}
\end{equation}
and the number density of a massless bosonic degree of freedom is
\begin{equation}
n_0 = \frac{\zeta(3)}{\pi^2} T^3 \ .
\end{equation}

For UED theories,
\begin{equation}
g_*(T) = g_*^{KK} D_d(T) \ ,
\end{equation}
Here $g_*^{KK}$ is the effective number of degrees of freedom per
KK level, and is a model-dependent constant. In all models,
however, there are more degrees of freedom at excited KK levels
than at the zero mode level, where many degrees of freedom are
projected out. For the SM, the effective number of degrees of
freedom, organized by spin, is $g_*^{\text{SM}} =
g_{0}^{\text{SM}} + \frac{7}{8} g_{1/2}^{\text{SM}} +
g_{1}^{\text{SM}} = 4 + \frac{7}{8} 90 + 24 = 106.75$.  For the
$S^1/Z_2$ UED model, at each KK level $n>0$ a gauge boson field
has 3 degrees of freedom and a fermion field has 4 degrees of
freedom, so the total effective number of degrees of freedom is
$g_*^{\text{KK}} = g_{0}^{\text{KK}} + \frac{7}{8}
g_{1/2}^{\text{SM}} \frac{4}{2} + g_{1}^{\text{SM}} \frac{3}{2} =
197.5$.  A similar counting may be done for other UED models once
the number of additional dimensions and the orbifold is specified.

The function $D_d(T)$ counts the number of excited modes in the
thermal bath at temperature $T$. $D_d(T)$ may be approximated by
counting
 all modes with mass below $T$. This yields, for $d$ extra dimensions,
\begin{equation}
D_d(T) = \frac{1}{2^d} V_d \left[ \frac{T}{\mKK} \right] ^d \ ,
\label{D}
\end{equation}
where
\begin{equation}
V_d = \frac{\pi^{d/2}}{\Gamma \left(1+\frac{d}{2}\right)} = 2, \pi,
\frac{4}{3} \pi , \frac{1}{2} \pi^2, \dots \ ,
\end{equation}
for $d=1,2,3,4,\dots$, is the volume of a unit spherical ball in $d$
dimensions, and the factor of $1/2^d$ in \eqref{D} accounts for the
restriction to $\vec{n}$ with non-negative components.

The number density of KK gravitons at level $\vec{n}$,
$n_{G^{\vec{n}}}$, is determined by the Boltzmann equation
\begin{equation}
\frac{d n_{G^{\vec{n}}}}{d t}+3 H n_{G^{\vec{n}}}
= C_{G^{\vec{n}}} \ ,
\label{boltzmann}
\end{equation}
where
\begin{equation}
C_{G^{\vec{n}}} = \sum
\langle \sigma_{\phi_a^{\vec{k}} \phi_b^{\vec{l}}
\to \phi_c^{\vec{m}} G^{\vec{n}}}  v \rangle n_{\phi_a^{\vec{k}}}
n_{\phi_b^{\vec{l}}}
\end{equation}
is the collision operator.  All graviton destruction processes are
negligible, given the low abundances and decay rates of gravitons.  We
will parameterize the collision operator as
\begin{equation}
C_{G^{\vec{n}}} = C \sigma \left[ g_*(T) n_0 \right] ^2 \ ,
\end{equation}
where
\begin{equation}
\sigma = \frac{\alpha_3}{4 \pi M_4^2} \ .
\end{equation}
Here $\alpha_3$ is the strong coupling constant and $C$ is a
constant. If every light degree of freedom could interact with every
other light degree of freedom with cross section $\sigma$,
$C_{G^{\vec{n}}}$ with $C = 1$ would be a reasonable estimate.
However, global and local symmetries restrict which reactions are
possible, and not all interactions involve strongly-interacting SM
particles.  A detailed calculation of $C$ can be done once a specific
UED model is chosen and its spectrum determined.  Here we keep the
analysis general by leaving $C$ as a free parameter.  Based on the
results of detailed studies of gravitino abundances from reheating in
supersymmetric models~\cite{Bolz:1998ek}, we expect values of $C \sim
{\cal O} (0.01)$, and we will consider a range of $0.001 \alt C \alt
0.1$ in present numerical results below. Given the high power
dependence of the graviton abundance on $\TRH$, we will see that
bounds on $\TRH$ are rather insensitive to even this generous range
for $C$.  Note also that, as $C=1$ is certainly an overestimate, if a
scenario passes all constraints even with $C=1$, it is certainly
allowed.

With these definitions, the KK graviton number density satisfies
\begin{equation}
\frac{d n_{G^{\vec{n}}}}{dt}+3 H n_{G^{\vec{n}}}
= C \sigma \left[ g_*(T) n_0 \right] ^2 \ .
\end{equation}
This is most conveniently solved by changing variables
$n_{G^{\vec{n}}} \to Y_{G^{\vec{n}}} \equiv n_{G^{\vec{n}}}/s$ and $t
\to T$.  Adiabaticity implies that entropy $S = s R^3 \propto g_*(T)
T^3 R^3 \propto T^{3+d} R^3$ is conserved, and so
\begin{equation}
\frac{1}{s} \frac{ds}{dt} = -3 \frac{1}{R} \frac{dR}{dt} = -3 H
\ , \qquad \frac{dT}{dt} = - \frac{3}{3+d} HT \ .
\label{change}
\end{equation}
With the relations of \eqref{change}, the Boltzmann equation becomes
\begin{equation}
\frac{dY_{G^{\vec{n}}}}{dT} = - \frac{3+d}{3} \frac{1}{HTs} \,
C \sigma \left[ g_*(T) n_0 \right] ^2 \ .
\end{equation}
$Y_{G^{\vec{n}}}$ changes until $G^{\vec{n}}$ production stops at
temperature $T \sim m_n$ and then remains constant until gravitons
begin to decay.  After BBN and before KK gravitons decay, then,
\begin{eqnarray}
Y_{G^{\vec{n}}} &=& \int_{m_n}^{\TRH}
dT \, \frac{3+d}{3} \frac{1}{HTs} \,
C \sigma \left[ g_*(T) n_0 \right] ^2 \nonumber \\
&=& \frac{45 \sqrt{5} \zeta^2(3)}{2 \pi^8}
\alpha_3 \frac{\mKK}{M_4} C \sqrt{g_*^{KK}} \,
\frac{3+d}{2+d} \, \sqrt{\frac{V_d}{2^d}}
\left[ \left( \frac{\TRH}{\mKK} \right)^{1+\frac{d}{2}}
- \left| \vec{n} \right|^{1+\frac{d}{2}} \right] \ .
\end{eqnarray}

For comparison with cosmological constraints, it is convenient to
determine the graviton energy density, normalized to the background
photon number density $n_{\gamma} = 2 n_0$, at the time of BBN:
\begin{equation}
\zeta_{G^{\vec{n}}}
\equiv \left.
\frac{ m_n n_{G^{\vec{n}}}}{n_{\gamma}}
 \right|_{\text{BBN}}
= m_n Y_{G^{\vec{n}}} \left. \frac{s}{n_{\gamma}}
\right|_{\text{BBN}}
= \frac{\pi^4}{45 \zeta(3)} g_*^{\text{BBN}}
\left| \vec{n} \right| \mKK Y_{G^{\vec{n}}} \ .
\end{equation}
The total graviton energy density is then determined by summing over
$\vec{n}$.  For $T$ significantly larger than $\mKK$, we may take the
continuum limit
\begin{equation}
\sum_{\vec{n}} \to \int d^dn = \int \frac{1}{2^d} A_d n^{d-1} dn \ ,
\label{vec}
\end{equation}
where
\begin{equation}
A_d = \frac{2 \pi^{d/2}}{\Gamma \left( \frac{d}{2} \right)}
= 2, 2\pi, 4\pi, 2\pi^2, \dots \ ,
\end{equation}
for $d=1,2,3,4,\dots$, is the area of a unit sphere in $d$ dimensions,
and the factor of $1/2^d$ in \eqref{vec} again accounts for the
restriction to $\vec{n}$ with non-negative components.  Integrating
over all KK levels up to $m_n = \TRH$, the total energy
density in KK gravitons is
\begin{eqnarray}
\zeta_G &=& \sum_{\vec{n}} \zeta_{G^{\vec{n}}}
\approx \int_0^{\frac{T}{\mKK}} \frac{1}{2^d} A_d n^{d-1}
\zeta_{G^{\vec{n}}} \, dn \nonumber \\
&=& \frac{\sqrt{5} \zeta(3)}{2 \pi^4} \alpha_3 \frac{\mKK^2}{M_4}
C \sqrt{g_*^{KK}} g_*^{\text{BBN}} \,
\frac{3+d}{(1+d)(4+3d)} \, \sqrt{\frac{V_d A_d^2}{2^{3d}}}
\left( \frac{\TRH}{\mKK} \right)^{2+\frac{3d}{2}}
\ . \label{zeta}
\end{eqnarray}
If KK gravitons decay to SM particles, $\zeta_G$ gives the total
energy density, normalized to the background photon density, deposited
during the era of graviton decay.

Alternatively, one may take the opposite limit, and assume that KK
gravitons decay predominantly through KK number preserving
interactions $G^n \to n \, \text{LKP} + X$, and nearly all of the
energy stored in KK gravitons exists now in the form of KK dark
matter.  In this scenario, the current dark matter energy density,
normalized to the critical density, is
\begin{equation}
\Omega_G = \zeta_G \frac{n_{\gamma}^0}{\rho_c^0}
\simeq \frac{\zeta_G}{13~\ev} \ ,
\label{Omega}
\end{equation}
where $n_{\gamma}^0 \simeq 410 ~\cm^{-3}$ is the present photon number
density, and $\rho_c^0 \simeq 5300~\ev~\cm^{-3}$ is the critical
density.

Numerically, given $g_*^{\text{BBN}} \simeq 3.36$ and $\alpha_3
\approx 0.1$, we find
\begin{eqnarray}
D=5\ : \quad
\zeta_G &=& 1.1 \times 10^{-14}~\gev \times
C \, \Biggl[ \frac{g_*^{KK}}{200} \Biggr]^{\frac{1}{2}}
  \Biggl[ \frac{\mKK}{1~\tev} \Biggr]^2
  \Biggl[ \frac{\TRH}{\mKK} \Biggr]^{\frac{7}{2}}
\label{numbegin} \\
\Omega_G &=& 8.4 \times 10^{-7} \times
C \, \Biggl[ \frac{g_*^{KK}}{200} \Biggr]^{\frac{1}{2}}
  \Biggl[ \frac{\mKK}{1~\tev} \Biggr]^2
  \Biggl[ \frac{\TRH}{\mKK} \Biggr]^{\frac{7}{2}} \ , \\
D=6\ : \quad
\zeta_G &=& 8.9 \times 10^{-15}~\gev \times
C \, \Biggl[ \frac{g_*^{KK}}{200} \Biggr]^{\frac{1}{2}}
  \Biggl[ \frac{\mKK}{1~\tev} \Biggr]^2
  \Biggl[ \frac{\TRH}{\mKK} \Biggr]^{5} \\
\Omega_G &=& 6.8 \times 10^{-7} \times
C \, \Biggl[ \frac{g_*^{KK}}{200} \Biggr]^{\frac{1}{2}}
  \Biggl[ \frac{\mKK}{1~\tev} \Biggr]^2
  \Biggl[ \frac{\TRH}{\mKK} \Biggr]^{5} \ , \\
D=7\ : \quad
\zeta_G &=& 5.0 \times 10^{-15}~\gev \times
C \, \Biggl[ \frac{g_*^{KK}}{200} \Biggr]^{\frac{1}{2}}
  \Biggl[ \frac{\mKK}{1~\tev} \Biggr]^2
  \Biggl[ \frac{\TRH}{\mKK} \Biggr]^{\frac{13}{2}} \\
\Omega_G &=& 3.9 \times 10^{-7} \times
C \, \Biggl[ \frac{g_*^{KK}}{200} \Biggr]^{\frac{1}{2}}
  \Biggl[ \frac{\mKK}{1~\tev} \Biggr]^2
  \Biggl[ \frac{\TRH}{\mKK} \Biggr]^{\frac{13}{2}} \ .
\label{numend}
\end{eqnarray}
Note that for each $D$, these expressions, derived from
\eqsref{zeta}{Omega}, are not mutually consistent --- the energy
density in gravitons is either deposited in SM particles, or in KK
dark matter (WIMP or superWIMP), or a mixture.  These expressions are
the maximal energy deposited in SM particles, and the maximal
primordial energy density of KK dark matter.  Which, if either, is
realized depends on the specific UED model, and detailed
considerations of higher KK mode cascade decays.  On the other hand,
these expressions are useful, as, if both are within existing bounds,
the model is guaranteed to be consistent with current constraints.

\section{Cosmological Constraints}
\label{sec:cosmology}

There are two sets of constraints on the KK graviton abundance.
First, gravitons $G^{\vec{n}}$ may decay to stable LKPs, such as $G^1$
or $B^1$.  The fraction of initial energy that winds up in LKPs is
highly dependent on the UED model and the spectrum of KK modes, which
determines the form of cascade decays.  Clearly the energy density in
LKPs cannot exceed the energy density of the initial KK gravitons.  We
may therefore impose the constraint
\begin{equation}
\Omega_G < \Omega_{\text{DM}} \approx 0.23 \ .
\end{equation}
If this constraint is satisfied, KK dark matter will not overclose the
universe, irrespective of details of the UED model.

The second set of constraints follows from requiring that SM particles
produced in late decays of KK gravitons not destroy the successful
predictions of BBN.  These constraints are complicated, depending
sensitively on the kind of energy deposited and the time at which it
is released.  For electromagnetic decays, the constraint on
$\zeta_G^{\text{EM}} = B^{\text{EM}} \zeta_G$ has been studied in
detail~\cite{Holtmann:1998gd,Jedamzik:1999di,Kawasaki:2000qr}, most
recently in Ref.~\cite{Cyburt:2002uv}.  These constraints are very
weak for decay times $\tau < 10^5~\s$, increase in stringency from
$\zeta_G^{\text{EM}} \alt 10^{-9}~\gev$ at $\tau \sim 10^6~\s$ to
$\zeta_G^{\text{EM}} \alt 10^{-12}~\gev$ at $\tau \sim 10^9~\s$, and
then remain roughly constant up to $\tau \sim 10^{12}~\s$.

For hadronic cascades, the picture is at present less clear.  For
$\tau \alt 10^2~\s$, hadronic constraints are relatively weak, but
they become $\zeta_G^{\text{had}} = B^{\text{had}} \zeta_G \alt
10^{-12}~\gev$ for $10^2~\s \alt \tau \alt
10^4~\s$~\cite{Kohri:2001jx}.  For longer lifetimes, there are no
detailed recent analyses.  The constraint may become weaker, but we
assume conservatively that it remains at the $\zeta_G^{\text{had}}
\alt 10^{-12}~\gev$ level.

{}From Eqs.~(\ref{numbegin})--(\ref{numend}), we see that both the
overclosure and BBN constraints may be satisfied for any $D$ and $\TRH
\sim \mKK \sim 1~\tev$, even for $C \sim 1$ and taking the extreme
cases in which all energy is deposited in either stable LKPs, EM
cascades, or hadronic cascades. Although there are many uncertainties
in this analysis, there is surely an allowed window in which the KK
graviton abundance satisfies all existing constraints.  This is
necessary to establish the viability of KK WIMP and superWIMP dark
matter.  For LKPs to achieve the desired thermal relic density, either
in the form of WIMPs such as $B^1$ or superWIMPs such as $G^1$, the
universe must be reheated to a temperature $\TRH \agt \mKK/25$. The
region with ideal KK WIMP thermal abundance~\cite{Servant:2002aq} is
also shown.  This result shows that there are consistent cosmologies
in which the reheat temperature is low enough to suppress the
primordial KK graviton abundance appropriately, but high enough to
generate the desired thermal relic abundance for WIMP or superWIMP
dark matter.

\begin{figure}[tbp]
\postscript{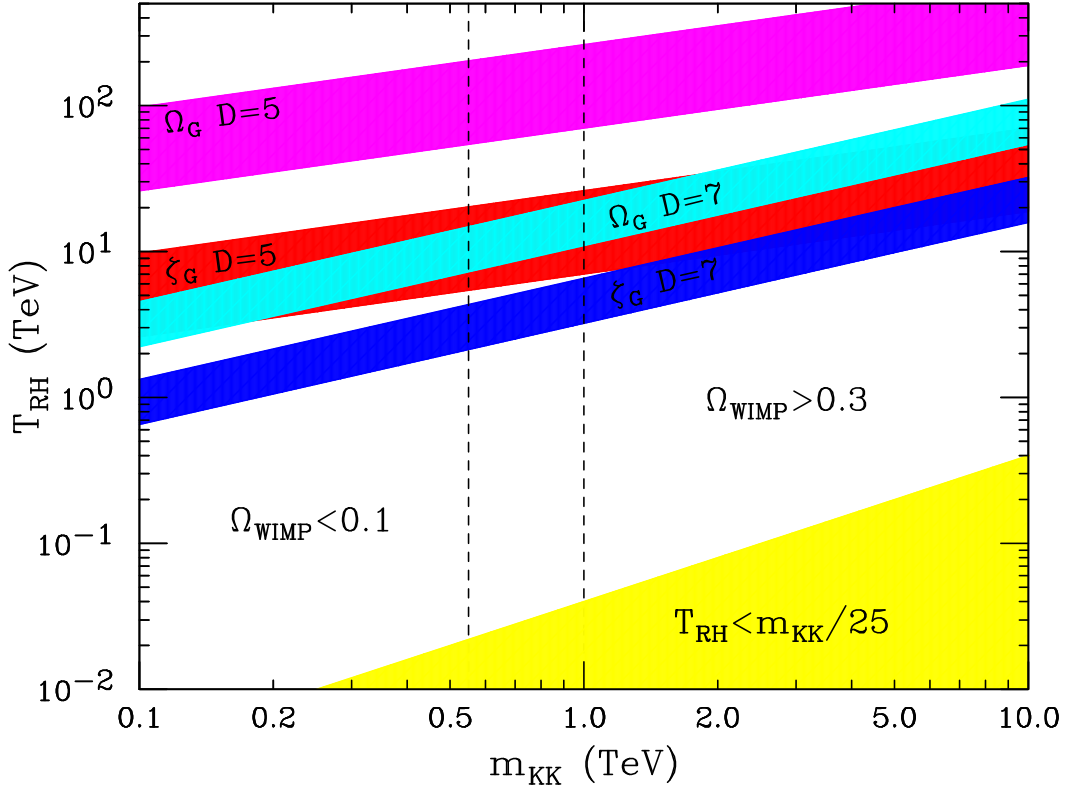}{0.85}
\caption{Bounds on the reheat temperature $\TRH$ as a function of
$\mKK$ from the overclosure constraint $\Omega_G < 0.23$ and the BBN
constraint $\zeta_G < 10^{-12}~\gev$ for $D=5,7$, as indicated.  We
assume $g_*^{KK} = 200$; the range in each bound arises from varying
$C$ from $0.001$ to $0.1$ (see text). In the region with $\TRH <
\mKK/25$, $\TRH$ is too low to generate the thermal relic abundance
for WIMPs.  The vertical bands delimit regions where the $B^1$ thermal
relic abundance is too low ($\Omega_{\text{WIMP}} < 0.1$),
approximately right, and too high ($\Omega_{\text{WIMP}} > 0.3$).
\label{fig:reheat} }
\end{figure}

At the same time, given the rather strong power law dependence of the
graviton energy density, overclosure and BBN stringently constraint
$\TRH$.  In Fig.~\ref{fig:reheat}, we plot the bounds on $\TRH$ as
functions of $\mKK$ for $D=5,7$ from constraints on overabundance and
energy release.  Roughly we find constraints $\TRH \alt 1 - 10~\tev$
for $100~\gev < \mKK < 1~\tev$.  In a given model, one of the
constraints may be inapplicable, but they cannot both be completely
avoided.  The range in each constraint results from varying $C$ in the
range $0.001 < C < 0.1$.  Note that the $\TRH$ limits are rather
insensitive to the substantial uncertainties encoded in $C$, given the
extreme sensitivity of graviton abundances to $\TRH$; for $C$ varying
by two orders of magnitude, the bounds on $\TRH$ vary by only factors
of 2 to 4, depending on the number of extra dimensions.

\section{Summary}
\label{sec:summary}

We have provided a general formalism for analyzing the dynamics of
gravitons in UED theories. In particular, we found the couplings of KK
gravitons to fermions and gauge bosons and presented the widths for
decays of excited fermions and gauge bosons into KK gravitons in
\eqsref{B1lifetime}{f1lifetime}.  These results are of special
relevance when a KK graviton is the LKP and a superWIMP candidate, as
they determine the observable implications of KK graviton dark matter
for, for example, BBN, the cosmic microwave background, and the
diffuse photon flux.

We have also determined the abundance of KK gravitons produced after
reheating in a general manner applicable to UED models for arbitrary
numbers of extra dimensions, and also more generally to other models
of extra dimensions.  The possibility of populating a large number of
graviton states at different KK levels implies that the production of
gravitons after reheating is extremely efficient and extremely
sensitive to the reheat temperature $\TRH$.  For $d$ extra dimensions,
the energy density in gravitons, given in \eqref{zeta}, is
\begin{equation}
\zeta_G = \sum_{\vec{n}} m_{G^{\vec{n}}} \frac{n_{G^{\vec{n}}}}{s}
\sim \frac{\mKK^2}{M_4} \left( \frac{\TRH}{\mKK} \right)^{2 +
  \frac{3d}{2}} \ .
\end{equation}
This is to be contrasted with the case of
gravitinos~\cite{Pagels:ke,Bolz:1998ek}, for which
\begin{equation}
\zeta_{\tilde{G}} = m_{\tilde{G}} \frac{n_{\tilde{G}}}{s}
\sim \frac{\mSUSY^2}{M_4}\, \frac{\TRH}{\mSUSY} \ .
\end{equation}
The constraints on $\TRH$ are therefore extremely stringent.  They are
presented in Fig.~\ref{fig:reheat} and are of the order of $\TRH \alt
1 - 10~\tev$ for $100~\gev < \mKK < 1~\tev$.

The constraints derived here are robust, being independent of the
gravi-scalar mass~\cite{Perivolaropoulos:2002pn} and applicable
irrespective of which KK particle is the LKP.  These constraints also
apply in the presence of KK-parity violating interactions, as these
will only serve to increase the primordial graviton production and
lead to the decay of gravitons to SM particles.  They supplement the
requirement that the reheat temperature be below the cutoff of the 4D
effective theory~\cite{Chivukula:2003kq} and rather severely constrain
ideas for leptogenesis.  Such low reheat temperatures also constrain
inflation scenarios, requiring, for example, that inflaton decay to SM
particles be suppressed by extremely small couplings or kinematically
through enhanced SM plasma masses at high
temperatures~\cite{Kolb:2003ke}.

At the same time, we have found that there exists a range of reheat
temperature with $\TRH \sim \mKK$ such that the primordial production
of KK gravitons is within bounds, but a thermal relic density of WIMPs
may be produced.  KK WIMP and superWIMP candidates are therefore
viable, despite the stringent graviton constraints applicable to these
extra dimensional theories.

\begin{acknowledgments}
We are grateful to T.~Han for helpful correspondence.
\end{acknowledgments}


\end{document}